\documentclass[11pt,twocolumn,conference,letterpaper,oneside]{IEEEtran}

\usepackage{epsfig,color,citesort}
\usepackage{subfigure}
\usepackage{times}
\usepackage{fancyhdr} % for including page numbers
\usepackage{lastpage} % just for Lastpage feature

\renewcommand{\figurename}{Figure}
\renewcommand{\refname}{}

\renewcommand{\baselinestretch}{1.0}

\newcommand{\squeezeFigureT}{\vspace{-0.4\baselineskip}}
\newcommand{\squeezeFigureC}{\vspace{-1.5\baselineskip}}
\newcommand{\squeezeFigureB}{\vspace{-0.8\baselineskip}}
\newcommand{\squeezeSection}{\vspace{0.0cm}}%{-0.2cm}}
\newcommand{\squeezesubSection}{\vspace{-0.1cm}}

\begin{document}

\chead{
 This work has been submitted to the IEEE milcom 2009 conference for possible publication. Copyright may be transferred without notice, after which this version may no longer be accessible.
 }

\title{	\fontfamily{cmr} \fontseries{b} \fontsize{12}{-10} \selectfont EFFICIENT AND PORTABLE SDR WAVEFORM DEVELOPMENT: THE NUCLEUS CONCEPT }
\bstctlcite{IEEEexample:BSTcontrol}
\author{\IEEEauthorblockN{ Venkatesh Ramakrishnan, Ernst M. Witte, \\ Torsten Kempf, David Kammler, \\Gerd Ascheid, Rainer Leupers, Heinrich Meyr}
        \IEEEauthorblockA{Institute for Integrated Signal Processing Systems,\\
                          RWTH Aachen University, Germany\\
                          ramakrishnan@iss.rwth-aachen.de}
                         \and
        \IEEEauthorblockN{}
        \IEEEauthorblockA{~~~and}
                          \and
        \IEEEauthorblockN{Marc Adrat, Markus Antweiler}
        \IEEEauthorblockA{Research Establishment for Applied Science (FGAN),\\
                          Dept. FKIE/KOM, Wachtberg, Germany\\
                         adrat@fgan.de}
        }

\maketitle

\thispagestyle{fancy}

\cfoot{\fontfamily{cmr} \fontsize{9}{0} \selectfont {~~~~~~~\thepage\ of \pageref{LastPage}}}
\begin{figure}[b]
\vspace*{-3.01ex}{\footnotesize \hfill This research project was
performed under contract with the \emph{Technical\\[-1ex] Center for
Information Technology and Electronics} (WTD-81), Germany.}
\end{figure}

%\begin{abstract}
%\normalfont{
%\emph{
%\section{Abstract}
\begin{center}
{\fontfamily{cmr} \fontseries{b} \fontsize{11}{-10} \selectfont ABSTRACT}
%\textbf{ \normalsize{ ABSTRACT}}
\end{center}
\emph{ Future wireless communication systems should be flexible to support different waveforms (WFs) and be cognitive to sense the environment and tune themselves. This has lead to tremendous interest in software defined radios (SDRs). Constraints like throughput, latency and low energy demand high implementation efficiency. The tradeoff of going for a highly
efficient implementation is the increase of porting effort to a new hardware (HW) platform. In this paper, we propose a novel concept for WF development, the Nucleus concept, that exploits the common structure in various wireless signal processing algorithms and provides a way for efficient and portable implementation. Tool assisted WF mapping and exploration is done efficiently by propagating the implementation and interface properties of Nuclei. The Nucleus concept aims at providing software flexibility with high level programmability, but at the same time limiting HW flexibility to maximize area and energy efficiency.}

%The definitions that are used in the concept are introduced and then explained with an example. The methodology for WF development using the nucleus concept is presented followed by the challenges that exist in the proposed concept. We believe that the concept will help in developing portable and efficient WFs for future radios.
%}
%\end{abstract}

\section{\fontfamily{cmr} \fontseries{b} \fontsize{11}{-10} \selectfont INTRODUCTION}
Flexibility in modern radio devices has been proposed to serve different goals, such as efficient multi-modal or multi-standard
transmission in order to support compliance with new and old WFs, promote interoperability and reduction of costs via
modular and parametric design. Furthermore, system cognition results in high flexibility requirements, which need to be efficiently
addressed by the transceiver leading to a solution like SDR system.

One of the key requirements for flexibility is WF portability across various HW platforms. Portability cannot be defined as a binary term but as an inverse to the porting effort \cite{Witte2008}. Portability can offer several key advantages like implementation reuse and fast time-to-market. It is also a goal of the joint
tactical radio system (JTRS) program \cite{jtrs}. Pure software solutions for a SDR offer maximum portability. Due to low
implementation efficiency and tough throughput, latency constraints pure software radios are not yet feasible.
Therefore, currently some of the processing intensive components are still implemented in dedicated application-specific integrated circuits thereby limiting portability.

WF implementation in C/C++ is portable, at-least, across platforms with programmable processing elements (PEs). But the
implementation is not efficient enough to meet the realtime constraints. For example, assembly can be more efficient by
one order of magnitude than a C program \cite{Kempf2008}. WF implementation at such a low level needs tremendous porting effort. Furthermore, C/C++ is not able to provide adequate support needed for physical (PHY) layer signal processing e.g. fixed point types, circular buffers, etc. Therefore, one of the key challenges that needs to be addressed in SDR development is to enable WF portability and maintain implementation efficiency at the same time.

\begin{figure}[h]
\squeezeFigureT
\centering \epsfig{file=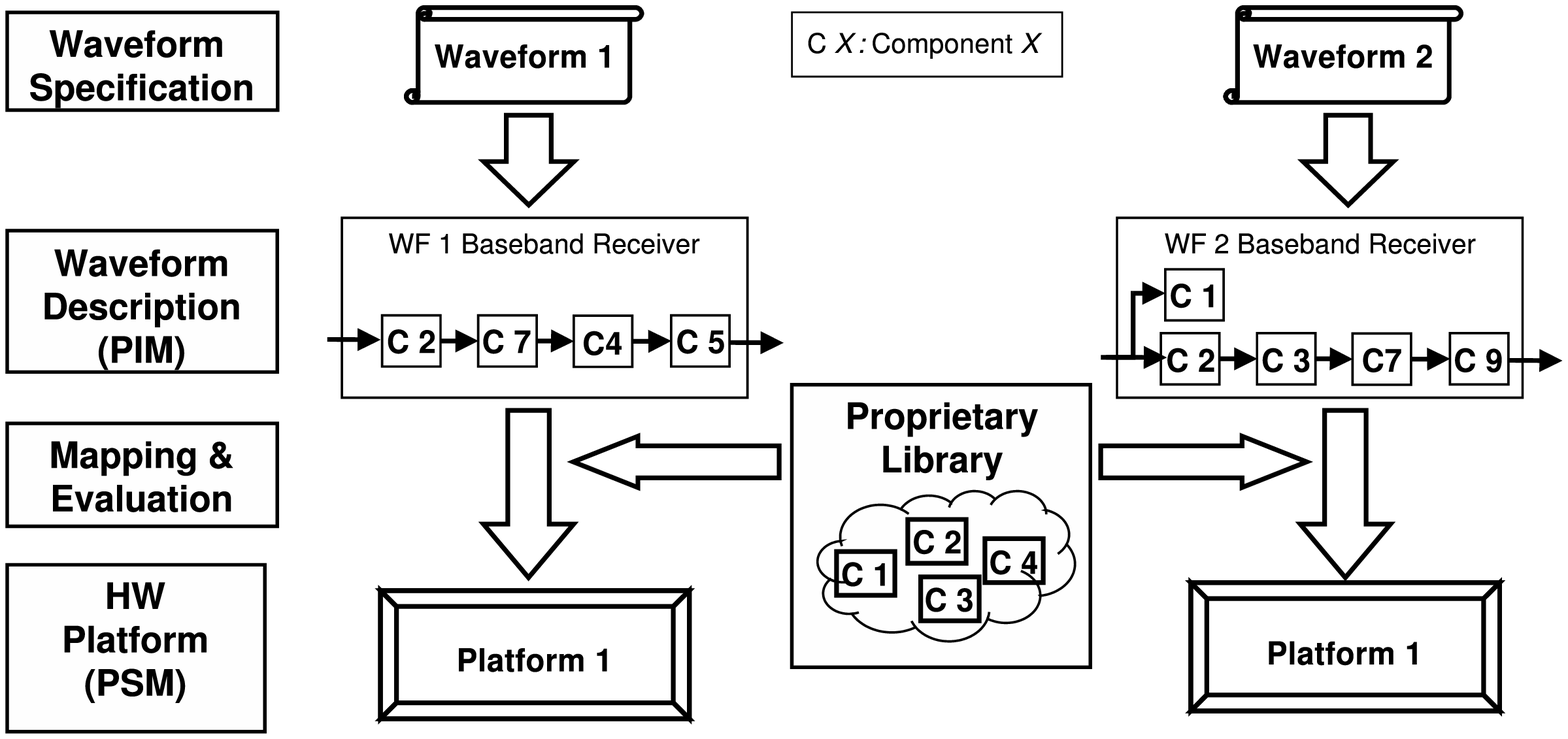, width=1.0\columnwidth}%
%\squeezeFigureC
\caption{Traditional Library Based Waveform Development} \label{fig:library_model} \squeezeFigureB
\end{figure}

This challenge is currently addressed by raising the abstraction level for WF implementations leading to library based approaches. As depicted in Figure~\ref{fig:library_model}, the basis of current approaches \cite{Lang2008,Glos2008,Blue2004,WimaxTI} is a library with basic components of a few WFs. A component \emph{X} is shown as \emph{CX} in Figure~\ref{fig:library_model}. The WF is constructed as a structured assembly of components, each of the components implementing a part of the WF functionality \cite{Pucker2006}.

The library provides efficient implementations of some basic components. From a platform independent model (PIM) of a WF a platform specific model (PSM) can be derived using the library. The advantage of this approach is that the same PIM can be used for porting to another HW. Also, it is flexible to develop components of different models independently. Library based approaches enable efficient utilization of heterogeneous multiprocessor system-on-chips (MPSoCs) due to the presence of hardware-aware implementations.

Currently, vendors maintain their own custom-built libraries for their platforms. Most of them are inaccessible to the public. There are no criteria for selecting components for the library which is crucial for a common library. We believe that the standardization of the library and its (components) interfaces is necessary and will lead to the development of a waveform description language (WDL). This will enable vendors not only to describe the WFs using this library but also to provide efficient implementations. Some important aspects which have to considered when building such a library are highlighted in the next section.

This paper proposes a new concept to enable WDL based WF development offering the following contributions

\begin{itemize}
\item An analysis is made on existing SDR WF development approaches to highlight portability and efficiency issues.
\item A new concept is proposed for tool assisted, portable and efficient WF development. Details on processes in a WDE are also presented.
\item Challenges in envisioning the concept are highlighted.
\end{itemize}

\squeezeSection
%\section{\textbf{Related Work and Motivation}}
\section{\fontfamily{cmr} \fontseries{b} \fontsize{11}{-10} \selectfont RELATED WORK AND MOTIVATION}
\label{sec:related}

Various solutions have been existing for developing SDRs. Solutions differ in portability, efficiency, usage of WDL, component granularity and reusability. This section provides an overview of some of the solutions and motivates the need for our Nucleus concept.

Many library approaches that exist for WF development are component based and/or model driven
\cite{Lang2008,Glos2008,Blue2004,WimaxTI}. As illustrated in Figure~\ref{fig:library_model}, some of the approaches \cite{Glos2008,Blue2004} have developed libraries for one particular HW platform (HW-specific). Others like \cite{WimaxTI,Lyrtech} have developed special libraries for one WF (WF-specific). Some library approaches are even both HW and WF specific. For example, the library in \cite{WimaxTI} has been
designed to support one type of WF, WiMax (with several modes), for one fixed HW using a TMS320TCI6482 DSP. The specifics of the above approaches limits porting of WFs. Since the components themselves and their interface are not standardized, implementations
from different vendors are not compatible in most scenarios and therefore lead to high integration effort. Furthermore, the
components of the library are functionality based (e.g. modulator/rake receiver). This results in limited commonality among WFs
leading to a reduced reuse of components and a huge implementation effort for a new WF. For example, when a WF uses a different modulation scheme whose component is not present in the library, the component has to be implemented. Such a functionality based library cannot be used for specifying a WF and will not lead to a WDL. A library of algorithmic kernels, using which the functionalities can be built, will improve usage among WFs.

The term WDL has been first introduced by E.D.Willink in \cite{Will2001}. The need for a WDL
arises from the specification-to-implementation problem. Specifying WFs using textual or mathematical/formal methods is not
efficient enough for automated implementation. A WDL aims at capturing the WF specification by a behavioral model and derive the implementation automatically. Though a WDL library is discussed, details about the library and portability issues are not provided.

In \cite{Gud2002}, Gudaitis presents a WDL concept based on the unified modeling language (UML), Matlab and the extensible markup language
(XML). The FM3TR WF has been used as an example to demonstrate the concept. Though language components of a WDL are presented,
information about the WDL library, reusability, realization and portability is not given.

The radio description language developed by Vanu Inc, targets WF portability  \cite{Chapin2001}. However, they have used C++ as
signal processing implementation language. It is also mentioned that few components (referred to as modules) were reused unchanged
from previously built WFs. But, information about the compute-intense (demanding) and the reused components is lacking. This is
necessary because highly demanding components need to be optimized to meet the WF requirements. The portability effort is highly
influenced by demanding components. From our investigations (\cite{Kempf2008,Witte2008}), implementation efficiency is low when a high level general-purpose language is used even with good compilers.

A methodology for selecting the components (referred to as common operators) for SDRs has been proposed in \cite{Moy2006}. Similar
to other existing library based approaches, the authors concentrate on identifying the common parts that exist in the implementation
of functionalities and not on the algorithms. Moreover, portability and mapping issues are not considered.

Few other important aspects have to be considered when targeting a library based approach enabling WDL. The data rate of
near-future mobile systems will be in the order of few hundred Mbits/sec requiring a processing power of hundreds of Giga operations per second. For such compute-intense systems, energy efficiency will be the key factor for the system design. HW platforms made up
only of PEs with general purpose architectures cannot provide the required efficiency. Heterogenous MPSoC platforms with multiple
processors having special architectures are better candidates with limited HW flexibility in order to gain efficiency.

In general, the selection of an algorithm has a huge impact on the performance. One algorithm that is good for one operational
scenario might not be good for another. Hence, it is necessary for the WF developer to have the flexibility of exploiting different
algorithms even with a fixed HW platform.

The granularity of the components is another important aspect that heavily influences re-usability and efficiency in a library based
WDL. If the components are too fine-grained, like a subtractor in \cite{Will2001}, it is inefficient to use them for constructing a
WF. If they are too coarse-grained, like a complete convolutional decoder in \cite{Chapin2001}, it limits reuse. Also, providing
optimized HW implementations based on such components is not efficient due to limited re-use. Therefore, the granularity should be between the two extremes representing a substantial part of WFs (critical) and the same time enabling re-use.

Even with a competent library, spatial and temporal mapping of a WF on a HW platform is a key factor that determines overall system efficiency. For an efficient mapping, the WF description should encode the information about its ideal mapping.  This information could be implementation properties like bit-width, type of scaling, etc. which can provide huge performance difference. The library and its components should be build such that it provides these properties using a WDL in an abstract manner. Tools in the waveform development environment (WDE) can use these properties to assist the developer for efficient mapping.

New SDR design approaches, in addition to being neither WF nor HW specific, should consider the above aspects during system design. Such approaches should enable (semi-) automatic generation of the implementation from the WF description that is not only efficient but also portable.

\squeezeSection
%\section{\textbf{The Nucleus Concept}}
\section{\fontfamily{cmr} \fontseries{b} \fontsize{11}{0} \selectfont THE NUCLEUS CONCEPT}
\label{sec:concept}

To overcome the drawbacks of traditional library based approaches, a new classification of library elements called Nuclei is
proposed targeting reusability, portability and implementation efficiency. Considering the important aspects for a library based WF
development, the Nucleus concept approaches system design by the following key principles:

\begin{itemize}
\item Limit HW flexibility to the minimum required level (for example, architecture of PEs, communications and memories)
\item Maximize area and energy efficiency
\item Manage/exploit flexibility by means of high level programmability
\end{itemize}

%This section describes the proposed concept. The concept is introduced along with the Definitions in section~\ref{sec:Define}. An example is given to explain the concept in section~\ref{sec:Example}. This is followed by the challenges that need to be addressed by the concept to realize the vision of WDL.

\squeezesubSection
\subsection{\fontfamily{cmr}\fontseries{b}\fontsize{11}{0}\selectfont DEFINITIONS}
\label{sec:Define} \squeezesubSection
\begin{itemize}
\item \emph{\textbf{Nucleus}} : A Nucleus is defined as a critical, demanding, flexible, algorithmic kernel that captures common functionalities within and/or among WFs.

\item \emph{\textbf{Genre}}: A Genre is defined as a set of algorithms containing the same Nucleus. An example for a Genre is illustrated in Figure~\ref{fig:nucleus_example}.

\item \emph{\textbf{Flavor}}: A Flavor is defined as an efficient and optimized implementation for one Nucleus. There can be several Flavors for one Nucleus. The Flavors can be based on several algorithms. Each or all of the Flavor(s) can
have tunable parameters.
\end{itemize}

\subsection{\fontfamily{cmr}\fontseries{b}\fontsize{11}{0}\selectfont METHODOLOGY}
\begin{figure}[h]
\squeezeFigureT
\centering \epsfig{file=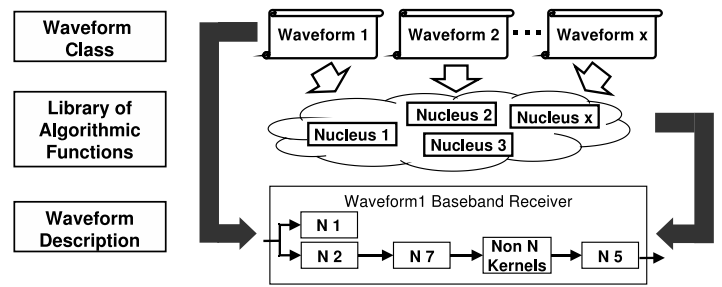, width=1.0\columnwidth}%
\vspace{-0.05cm} \caption{The Nucleus Concept} \label{fig:nucleus_concept} \squeezeFigureB
\end{figure}

As shown in Figure~\ref{fig:nucleus_concept}, the concept proposes to build a library of kernels from a class of WFs. A Nucleus kernel is shown as \emph{N} in the figure. These kernels are of algorithmic nature and do not need to represent any WF or implementation specific features. The library forms the basis for constructing a WF from the specification (Figure~\ref{fig:nucleus_concept}).

The key difference to the component-based WF development is that the Nuclei represent not only the functionality of signal
processing blocks on a different level of abstraction, but also their properties required for exploration. Properties that affect
the WF performance are essential for exploration. Therefore, information on the input data structure, bit-width, type of scaling, usage of truncation or rounding etc. will be abstracted and provided to the WF developer to aid exploration.

\begin{figure}[h]
\squeezeFigureT
\centering \epsfig{file=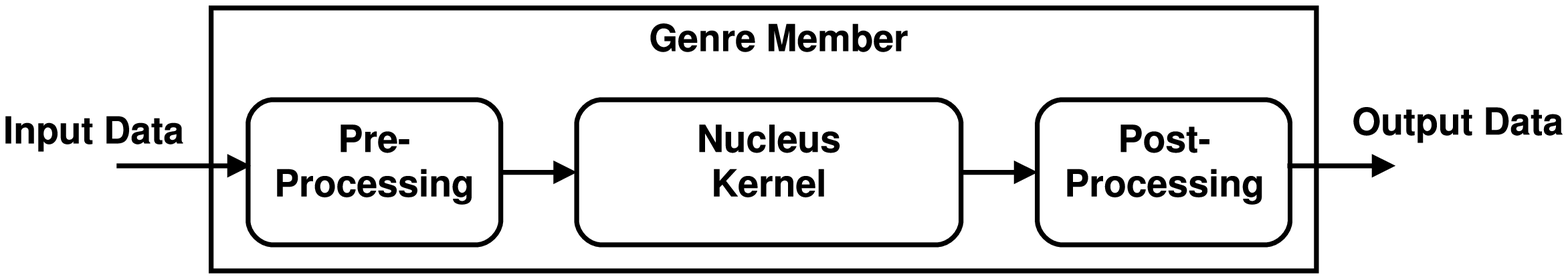, width=1.0\columnwidth}%
%\squeezeFigureC
\caption{Nucleus Genre Relationship} \label{fig:nucleus_genre} \squeezeFigureB
\end{figure}

A Nucleus does not necessarily represent functionality of parts of the WF directly (e.g. equalizer or demodulator). The
functionality is built using these kernels. Since the members of the Genre have the same computation and communication pattern,
they can be implemented using their Flavors. But, this might need some pre or post- processing in addition
(Figure~\ref{fig:nucleus_genre}). Since Flavors are optimized implementations, they can introduce additional constraints, e.g.
requiring a certain input data format like Q15. Part of the pre-processing in this case would be the adoption of the data format. Flavors need to be flexible enough to implement the extra processing in an efficient way.

One important aspect of the proposed approach is that different vendors can provide Flavors of some or all of the Nuclei in the form of a board-support-package (BSP). Flavors could be assembly codes for an
ASIP or DSP  and/or IP cores in HW or on a reconfigurable FPGA. Therefore, a Flavor in general can be seen as a
bundle of the PE and low-level software (if applicable). Due to this bundling it will be possible to capture the Nucleus
functionality by means of a HW independent high level Nucleus application programming interface (API). This enables to describe a WF
independently of the targeted MPSoC platform. Still, the WF can be mapped efficiently to the platform due to the bundled Nuclei
low-level implementation. Participation of different vendors is also possible due to the standardization of the functionality and
the interface of Nuclei. Vendors can also provide simulation models of the Nuclei e.g. in Matlab/Simulink.

The Nucleus concept enables WF developers to program the HW platform on a very high level (Nucleus level) as the Nuclei are visible via algorithm factors (Genre, Flavors). The developer will still be able to explore among the existing Flavors for WF
implementation. Since the programming is done at the Nucleus level hiding the underlying HW platform, it is efficient and simple for
the developers.

Even though new WF implementations go through many rounds of development and adjustment, the core algorithmic kernels often evolve
at a relatively slow rate \cite{Fan2009}. Since the proposed concept exploits such core common structures that exist in receiver
algorithms \cite{Meyr2007}, various WFs can be built even after the availability of the HW platform. Due to existence of such kernels in general purpose applications research in the same direction is also done by computer science experts \cite{Berkeley}.
\squeezesubSection
\subsection{\fontfamily{cmr}\fontseries{b}\fontsize{11}{0}\selectfont AN EXAMPLE}
\label{sec:Example} \squeezesubSection

FFT is an example of a Nucleus. FFT has been used for several decades in diverse applications. But the core algorithm itself has not evolved in the same pace as the implementations \cite{fftw}. The granularity is at an optimum level that enables reuse in various applications. We are using the existing work based on the FFT algorithm as an example for explaining the definitions in our concept. The references to the existing work are given as the explicit proof for our arguments. Since the motive is not to explain the example but to show the relationship between the definitions only brief comments are given for each Genre member.

\begin{figure}[h]
\squeezeFigureT
\centering \epsfig{file=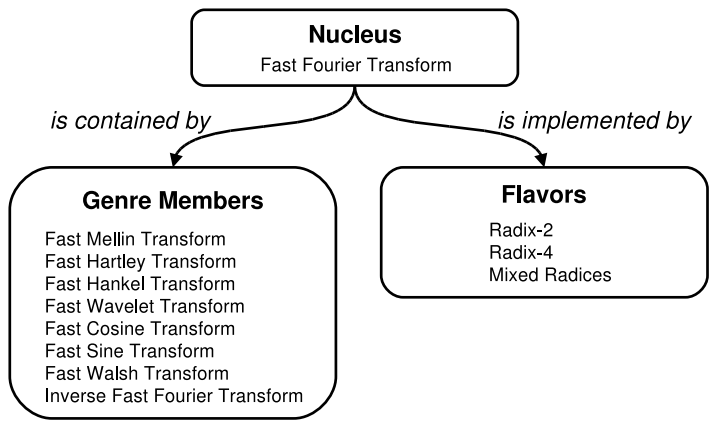, width=1.0\columnwidth} \squeezeFigureC
\caption{Nucleus Example}
\label{fig:nucleus_example} \vspace{-1.0\baselineskip} %\squeezeFigureB
\end{figure}

\begin{figure}[h]
\squeezeFigureT \centering \epsfig{file=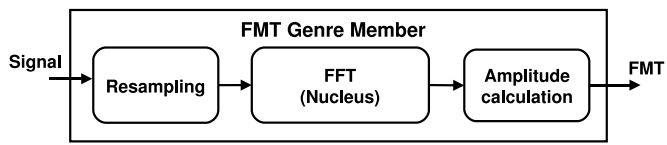, width=1.0\columnwidth} \squeezeFigureC \caption{Nucleus Genre Example} \label{fig:genre_flavor_example} \squeezeFigureB
\end{figure}

Figure~\ref{fig:nucleus_example} depicts the relation between the definitions that were introduced in the concept. As illustrated in the figure, for one Nucleus (e.g. FFT) there could be several members belonging to its Genre.  The Genre member fast Mellin transform (FMT) can be realized using the FFT kernel \cite{Chen1998}. But this requires pre and post-processing as shown in Figure~\ref{fig:genre_flavor_example}. Pre-processing in this case is re-sampling the input and post-processing is the amplitude calculation. Flavors need to implement this extra processing also efficiently. Fast Hankel transform (FHT) is the (one dimensional) Fourier transform of the Abel transform \cite{Hansen1985}. Here, the pre processing portion is the Abel transformation. Also, FHT using fast cosine transform and fast sine transform can be found in \cite{Knock2000}. FFT can be used as a basis for doing fast wavelet transforms \cite{Sava1997}. Discrete cosine transform and Walsh
transform can also be realized with pre and post-processing \cite{Tell2003}. Efficient implementations using an FFT kernel as basis
for realizing Walsh-Hadamard transform, discrete cosine transform (DCT) and discrete Hartley transform exist in \cite{Klo2002}. In
addition to that, the inverses of some of the above transforms belong to the same Nucleus, e.g. IFFT and IDCT.

Flavors for FFT have been existing using different algorithms and radices. For example, an FFT Flavor might use either a Cooley-Tukey
or Sandey-Tukey algorithm depending on the decomposition of FFT stages. Flavors can be available by using a single radix like
radix-2 or radix-4. If the number of FFT points is not a multiple of the basic radices, combination of different radices is present in one
Flavor (Figure~\ref{fig:nucleus_example}). The efficient implementations from the BSP could be assembly codes for an application specific instruction processor (ASIP)
or digital signal processor (DSP) (e.g. optimized FFT implementations from TI library \cite{TI2003}) and/or IP cores in HW or on a
reconfigurable field programmable gate array (FPGA) (e.g. FFT intellectual property (IP) cores from Xilinx \cite{Xilinx2008} and
Altera \cite{Altera2008}). The TI library \cite{TI2003} even provides support for IFFT. Simulation models for Flavors are also provided by vendors \cite{Xilinx2008,Altera2008,TI2003}. One Flavor might be
different to another in algorithm, flexibility and performance. For example, a Flavor with radix-2 has more FFT stages compared to radix-4.
Therefore, it is more flexible to scale the outputs in the intermittent stages than radix-4. Flavors can offer a wide performance
range with respect to latency and throughput, e.g. \cite{Xilinx2008}.

From the above references, it can be inferred that the members in the Genre can be realized using the same FFT kernel. As a
special case, an FFT kernel can be used for both PHY layer (e.g. OFDM) and application layer (e.g. video/audio compression) \cite{Tell2003} signal processing.

\squeezeSection
%\section{\textbf{WAVEFORM DEVELOPMENT ENVIRONMENT}}
\begin{center}
\section{\fontfamily{cmr} \fontseries{b} \fontsize{11}{10} \selectfont WAVEFORM DEVELOPMENT ENVIRONMENT CONCEPT}
\end{center}
\label{sec:wde-concept}

\begin{figure}[h]
%\squeezeFigureT
\centering \epsfig{file=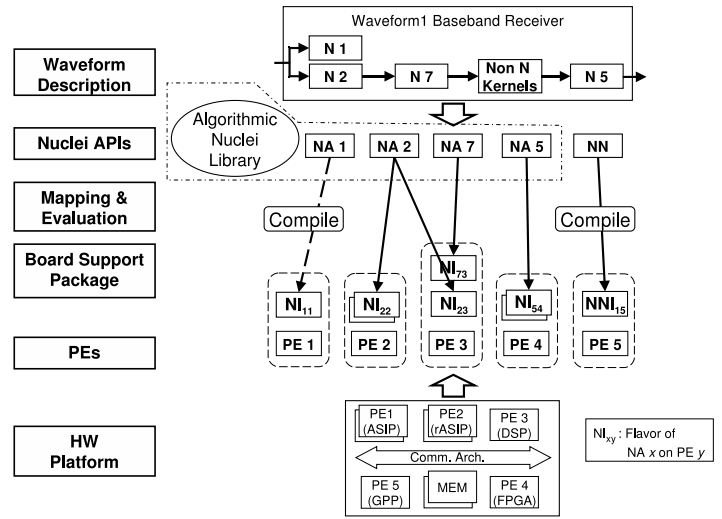, width=1.0\columnwidth} \squeezeFigureC \caption{Waveform Development Concept} \label{fig:wde}
\squeezeFigureB
\end{figure}

In general, a WDL forms the basis of a WDE. WDE denotes a system-level tool suite for automated WF development for SDRs. The tool-suite should cover the entire WF development process of requirements, design, implementation, integration and testing for SW and HW \cite{Gud2001}. Figure~\ref{fig:wde} depicts these processes with respect to our proposed concept.
While the WF description captures the requirements and design of the WF under consideration, implementation is provided by the BSP. Integration and testing processes are done in the mapping and evaluation stage (Figure~\ref{fig:wde}). NI represents the Nucleus implementation of one Nucleus kernel. The first subscript of NI denotes the Nucleus API (NA) and second
subscript denotes the PE for which the implementation is available in the BSP. Therefore, NI$_{23}$ denotes the NI of Nucleus 2 on PE 3.\\

\subsection{\fontfamily{cmr}\fontseries{b}\fontsize{11}{0}\selectfont WAVEFORM DESCRIPTION}
\squeezesubSection

The WF developer uses WDL for describing the WF using the API on the basis of the Nuclei library. The WF description will then consist of Nuclei kernels and non-nuclei (NN) kernels connected by communication links (Figure~\ref{fig:nucleus_concept}). The description also contains information about critical paths, constraints and control flow of the WF. The NN kernels reflect the other computation-light functionalities of the WF.

%\squeezesubSection
\subsection{\fontfamily{cmr}\fontseries{b}\fontsize{11}{0}\selectfont WAVEFORM IMPLEMENTATION}
%\squeezesubSection

The BSP plays an important role in the WF implementation. It provides efficient implementations of some or all of the Nuclei
kernels. This can enable tool based WF development where the Nuclei  kernels are replaced by the BSP and the rest of the WF (like NN
kernels) is built using a conventional approach (e.g. C/C++ code). If the implementation for a kernel that is in the WF
description is not available in the BSP, it also has to be built using the conventional approach. This is indicated by the dotted
line in Figure~\ref{fig:wde} for NA1. However, depending on performance  requirements, Nuclei kernels for which the BSP is missing
may have to be optimized for a given platform.

Since the software code of the NIs is highly optimized it achieves high performance, but it is bound to
that particular PE. As the input data structure of a Flavor will be known from the BSP, WDE can generate additional logic required for gluing Flavors.  This results in (semi-) automatic generation of WF implementation. Some of the features of WDE concept using the Nucleus approach are
\begin{itemize}
\item   The Flavors in different BSPs can use different algorithms to implement the same behavior while offering various tradeoffs.
\item   There can be several and different Flavors for one Nucleus in one BSP. This provides the flexibility to the WF designer to choose among the existing implementations according to the need.
\end{itemize}
\squeezesubSection
\subsection{\fontfamily{cmr}\fontseries{b}\fontsize{11}{0}\selectfont MAPPING}
\begin{figure}[h]
\squeezeFigureT \centering \epsfig{file=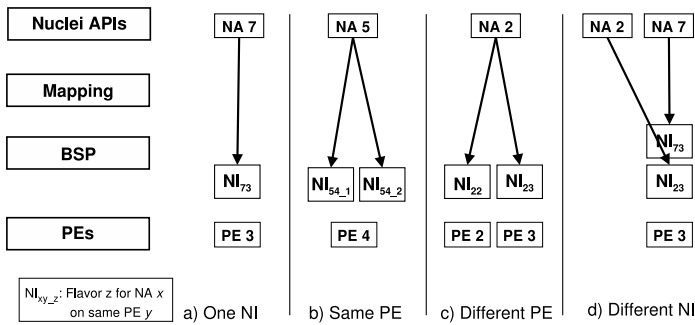, width=1.0\columnwidth} \squeezeFigureC \caption{Flavor Selection
Scenario} \label{fig:flavor_selection} \squeezeFigureB
\end{figure}

\squeezesubSection The WF description and the NI from the BSP form the basis of WF
mapping onto the HW platform. Even though Flavors are efficient, the overall system performance is not guaranteed and it is heavily
influenced by the spatial and temporal mapping. Figure~\ref{fig:wde} sketches the scenario of spatial mapping for a given HW
platform. The key difference to the traditional WF spatial mapping is that the kernel from the WF description is mapped not to the HW PE, but to the available NI from the BSP. This is possible due to the bundling of NI to the PE. As shown in Figure~\ref{fig:flavor_selection}, there might be several choices for selection
depending on the BSP.
\begin{itemize}
\item For one Nucleus, there can be only one NI (Figure~\ref{fig:flavor_selection}a).
\item For one Nucleus, there can be several NIs available for one PE (Figure~\ref{fig:flavor_selection}b).
\item For one Nucleus, there can be several NIs available for several PEs (Figure~\ref{fig:flavor_selection}c).
\item On one PE, there can be several NIs available for different Nuclei (Figure~\ref{fig:flavor_selection}d).
\item NIs can come from different vendors (Figure~\ref{fig:flavor_selection}d).
\end{itemize}

The presence of several mapping choices advocate the need for tool assistance. The choice of the NI will be based not only on the performance but also on the data structure of the neighboring NI interfaces. Since the NN kernels are computationally light, they might be mapped to the general purpose core (PE5) as shown in
Figure~\ref{fig:wde}.

One or more Nuclei (and/or non-nuclei) kernels can be mapped onto a single PE. This requires temporal mapping/scheduling of kernels
based on the data dependencies. Evaluation is needed to make sure that the constraints of the WF are met.
%\squeezesubSection
\subsection{\fontfamily{cmr}\fontseries{b}\fontsize{11}{0}\selectfont EVALUATION}
\squeezesubSection
A detailed evaluation of the mapping decisions is necessary primarily to check if the WF constraints are met. It
will also aid in finding a mapping that maximizes total system performance. The WF constraints need to be verified in terms of latency,
throughput, bit error rate and critical loops. To maximize total system performance, metrics are needed with which different
mapping decisions can be evaluated. The following metrics can be used:
%\vspace{-0.3\baselineskip

\begin{itemize}
\item Increasing data localization
\item Minimizing communication
\item Minimizing synchronization
\item Maximizing PE utilization
\item Maximizing energy efficiency
\end{itemize}

The mapping exploration should be done in an iterative manner and at different abstraction levels to trade-off between simulation
speed and accuracy. The WDE will provide the infrastructure necessary for doing the exploration and simulation for evaluating the
mapping decisions. The mapping challenge is to identify the right choice of NIs that are not only compatible to each other and meet
the requirements of the WF but also maximize the overall system efficiency. Since the mapping and evaluation will be an iterative process making it tool assisted can ease the whole process.

\squeezesubSection
\subsection{\fontfamily{cmr}\fontseries{b}\fontsize{11}{0}\selectfont ADVANTAGES}
\squeezesubSection The WDE methodology using the Nucleus concept delineates WF requirements in a manner that is neither application
specific nor HW specific. This leads to the following main advantages in realizing SDRs.
\begin{itemize}
\item \textbf{Portability}: Since the WF description is not HW specific, the same description can be ported to multiple HW platforms.
\item \textbf{Efficiency}: HW specific Flavors of the WF aids in efficient implementation.
\item \textbf{Reusability}: Since the approach is not WF specific the same NAs can be used for describing several WFs and this will enable WDL based radio development.
\item \textbf{Flexibility}: Since the interface and functionality of the Nuclei will be known through API, different vendors can provide different Flavors.
\item \textbf{Future Use}: This approach paves way for capturing future WFs.

%Some of the above advantages are also due to component based WF development \cite{Puck2007}. The important aspects of WDE has been discussed in this section. Our approach to realize the Nucleus concept will be presented in the next section.
\end{itemize}

\squeezeSection
\section{\fontfamily{cmr} \fontseries{b} \fontsize{11}{-10} \selectfont CHALLENGES}
\label{sec:nuclei-identification}
\squeezesubSection
%\subsection{CHALLENGES}
%\squeezesubSection

For realizing the Nucleus concept, it is necessary to explore the Nucleus design space in the algorithmic, HW and tools domain
in order to identify a basic set of Nuclei which will allow the realization of a wide range of WFs. Furthermore, it is
important to investigate the flexibility required for each NI to allow the usage for multiple WFs. All these aspects always need to consider energy efficiency. The challenges that exist to realize the Nucleus concept are described in this section.

\textbf{Nuclei Identification}\hspace{1mm} One big challenge is the identification of Nuclei kernels that exist among WFs.
The goal is to modularize or reformulate the algorithm in such a way that efficient implementation and reusability across multiple WFs is enabled. The energy efficiency of the Nucleus interconnect structure will have a significant influence on the whole SDR system. The interconnection issues are tightly coupled with the definition of a proper API for a Nucleus library and therefore it will provide good hints for WDL specification.

\textbf{Mapping}\hspace{1mm} Mapping a portable WF onto a HW platform spatially and temporally meeting the constraints is one of the key challenges. This requires exact knowledge of the HW interfaces of the NIs in order to deal with the required additional glue logic to connect the different HW blocks and manage their execution.

\textbf{Waveform Development Environment}\hspace{1mm} Providing easy-to-use programmability for complex receiver systems is
essential to exploit efficiently the system resources. A programming model that can bridge the gaps between WF, HW platform and
mapping is needed. A fast system simulation environment is key in order to validate mapping decisions. %The elements that are
%required for an efficient system-integration e.g. in terms of control flow, data flow, etc. needs to be investigated.
% The baseline of a set of NIs can be a working virtual or real prototype of a partial transceiver suitable to proof the objectives in both the areas of the WDE and the Nucleus exploration.

% \squeezesubSection
% \subsection{INTERACTION WITH OTHER LAYERS}
% \squeezesubSection

\textbf{Cross Layer Optimization}\hspace{1mm} For an efficient radio design it is mandatory to look at the PHY-media access control (MAC) interactions and enable cross-layer optimization. The focus should be not only on
the modularization and reuse of higher layer functionalities but also on the interfaces between the layers.

%\squeezesubSection
%\subsection{SCA COMPLIANCE}
%\squeezesubSection
\textbf{SCA Compliance}\hspace{1mm} Supporting software communications architecture (SCA) compliant WF development is key for implementing military WFs. The proposed concept is aligned with several aspects from the JTRS SCA community. For instance, the information that is fed into the mapping process of the WDE resembles the software assembly descriptor which describes the assembly of software components and HW devices. Furthermore, additional logic need to be generated to glue NAs to the SCA compliant APIs.  Though the propinquity of the Nucleus concept and JTRS SCA is visible in many ways, complete SCA compliancy must be guaranteed.

\squeezeSection
%\section{\textbf{Conclusions and Outlook}}
\section{\fontfamily{cmr} \fontseries{b} \fontsize{11}{-10} \selectfont CONCLUSIONS AND OUTLOOK}
\label{sec:outlook}

A novel concept for developing tool assisted, portable and efficient WFs for SDRs targeting a WDL has been proposed in this paper. By identifying the common, processing intensive, algorithmic kernels a Nuclei library is built. WF is described in a WDL using the Nucleus library. The optimal granularity of the kernels enable efficient implementations and reuse.  The standardization of the library will lead to the availability of Flavors from vendors. This provides algorithmic and implementation trade-offs even in a fixed HW platform. The implementation related properties is propagated to the Nucleus-API. Since the API has information for ideal mapping, efficient and tool assisted mapping can be provided by the WDE. Key challenges that exist in envisioning the concept were also presented in this paper.

Future work will address these challenges by working closely in algorithm, HW and tools domain. The selected application scenario for approaching and validating the Nucleus concept is an iterative and flexible MIMO-OFDM transceiver focusing on PHY layer, MAC layer and system cognition. A part of the transceiver will be implemented as a proof-of-concept. Investigations will also be done to find the properties that are needed for enabling (semi) automatic tool assisted WF development.

\squeezeSection
\begin{center}
\textbf{
\normalsize{
REFERENCES}}
\end{center}
\vspace*{-1.1cm}

\linespread{0.82}
\begin{small}
\bibliographystyle{IEEEtran}
\bibliography{milcom}
\end{small}

\end{document}